\documentclass[namedreferences]{solarphysics}

\usepackage[hyperref,optionalrh,showbiblabels]{spr-sola-addons} 
\usepackage{graphicx}        
\usepackage{amssymb}        
\usepackage{color}           
\usepackage{breakurl}        

\usepackage{ulem}
\usepackage{color}

\newcommand{\aap}{    {\it Astron. Astrophys.}}

\newcommand{\solphys}{{\it Solar Phys.}}

\chardef\us=`\_

\begin{document}

\begin{article}
\begin{opening}

\title{Combined Radio and Space-based Solar Observations: From Techniques to New Results - Preface}

\author[addressref={aff1},email={{eduard@eris.astro.gla.ac.uk}},corref]{\inits{E. P.}\fnm{Eduard P.}~\lnm{Kontar}\orcid{0000-0002-8078-0902}}
\author[addressref=aff2,email={{anindos@uoi.gr}}]{\inits{A.}\fnm{Alexander}~\lnm{Nindos}\orcid{0000-0003-0475-2886}}

\address[id=aff1]{School of Physics and Astronomy, University of Glasgow,
Glasgow, G12~8QQ, UK}
\address[id=aff2]{Physics Department, University of Ioannina, GR-45110 Ioannina, Greece}

\runningauthor{E.P. Kontar and A. Nindos}
\runningtitle{Combined Radio and Space-based Solar Observations}

\begin{abstract}
The phenomena observed at the Sun have a variety of unique radio
signatures that can be used to diagnose the processes in the solar
atmosphere.  The insights provided by radio
obervations are further enhanced when they are combined with
observations from space-based telescopes. This special issue
demonstrates the power of combination methodology at work and provides
new results on i)  type I solar radio bursts and thermal
emission to study active regions; ii) type II and IV bursts to
better understand the structure of coronal mass ejections;
iii)~non-thermal gyro-synchrotron and/or type III bursts to improve
characterization of particle acceleration in solar flares.  The
ongoing improvements in time, frequency, and spatial resolutions  of
ground-based telescopes reveal new levels of solar phenomena
complexity and pose new questions.
\end{abstract}
\end{opening}

CESRA, the Community of European Solar Radio
Astronomers\footnote{See webpage http://cesra.net.},
organizes triennial workshops on the investigations of the solar atmosphere
processes using radio and other observations.
The  2016 workshop\footnote{See http://cesra2016.sciencesconf.org.}
had a special emphasis on the complementarity of the current
and future space-based observations with the ground-based radio observations.
It was the place to discuss the new exciting science opportunities
that arise from the radio instruments like
the Atacama Large Millimeter/submillimeter Array (ALMA\footnote{See http://www.almaobservatory.org.}),
the Expanded Owens Valley Solar Array (EOVSA\footnote{See http://ovsa.njit.edu.}),
the Expanded Very Large Array (EVLA\footnote{See http://www.aoc.nrao.edu/evla/.}),
the Low Frequency array (LOFAR\footnote{See http://www.lofar.org.}),
the Mingantu Spectral Radioheliograph
\cite[MUSER, see][for details]{2009EM&P..104...97Y},
and the developments of the Square Kilometer Array (SKA\footnote{See https://www.skatelescope.org.}).
The workshop discussions have focussed on particle acceleration and transport,
the radio diagnostics of coronal mass ejections,
fine structures in solar radio bursts and the radio aspects of space weather
and this volume provides a snapshot of the developments 
and challenges discussed during the workshop. 
This topical issue covers four sub-topics:
\begin{enumerate}
  \item Solar Radio Emission Modelling \citep{Lyubchyk2017,Rodger2017,Stupishin2018,Zaitsev2017}
  \item Solar Flares and Solar Energetic Particles \citep{Anastasiadis2017,Benz2017,Altyntsev2017}
  \item Fine Structures in Solar Radio Emission \citep{Mohan2017,Mugundhan2017}
  \item Coronal Mass Ejections \citep{Al-Hamadani2017,Kumari2017,Long2017,Melnik2018,Miteva2018}
\end{enumerate}

\section*{Solar radio emission modelling}
 The complexity of the solar atmosphere as well as the increasing
 quality of solar radio observations necessitates
 the development of new more complex models.
The modeling of solar atmosphere parameters above sunspots
using RATAN-600 microwave observations
\citep{Stupishin2018}
inferred the upper transition-region structure of sunspots.
The method presented is based on iterative correction of the temperature–height profile
in the transition region and lower corona allowing  to test
time-independent models of density and temperature as a function of height.
Anticipating the future observations in the sub-THz range, \citet{Rodger2017}
studied how the ratio of brightness temperatures at two frequencies
can be used to estimate the optical thickness and the emission measure
for prominences. Highlighting that there is no generally accepted theory
explaining high brightness temperatures in type I storms,
\citet{Lyubchyk2017} proposed a new model to explain type I solar radio
bursts associated with active regions.
The model is based on the turbulence of kinetic-scale Alfvén waves
that produce an asymmetric plateau in the electron velocity
and a high level of Langmuir waves leading to plasma emission.
The model proposed by \citet{Zaitsev2017} suggests
that the electron acceleration and storage of energetic particles
in solar magnetic loops can be better explained by a mechanism
based on oscillations of the electric current. Specifically,
the model aims to explain synchronous pulsation in a wide frequency
interval that is hard to achieve by the sausage and kink MHD modes.

\section*{Solar flares and Solar Energetic Particles}
Solar flares are well-known for the efficient acceleration
of non-thermal electrons and hence being a source
of various solar radio bursts
\cite[see][for a review]{2008SoPh..253....3N}.
However, \citet{Benz2017} have demonstrated that there are exceptions to this rule
and presented observations of a radio-quiet solar pre-flare, which suggests
that acceleration to relativistic energies, if any, should be occurring with low efficiency
and does not lead to observable radio emission.
Studying optically thin gyrosynchrotron emission, \citet{Altyntsev2017}
reported an unusual flare, where the emission displays an apparently
ordinary mode polarization in contrast to the classical theory that
favors the extraordinary mode. This apparent ordinary wave emission
in the optically thin mode has been attributed to radio wave propagation
across the quasi-transverse layer that changes the radio-wave polarization.

Solar flares are often associated with the energetic particle events (SEPs)
observed near the Earth. These SEP events are an important element of space weather
and there is a growing interest in development of reliable forecasting systems.
\citet{Anastasiadis2017} presented an integrated prediction system
for solar flares and SEP events. The system is based on statistical methods and
demonstrated promising results for the expected SEP characteristics.

\section*{Fine Structures in Solar Radio Emission}
The high frequency resolution solar radio telescopes have enabled the detailed
imaging and spectroscopic studies of the fine structures of solar radio
emission \citep{2017NatCo...8.1515K}. For example, type III bursts that sometimes
show fine structures (stria) in dynamic spectra \citep{1972A&A....20...55D}
can be used to study density fluctuations.
Assuming that the individual stria bandwidth are determined
by the amplitude of density fluctuations,
\citet{Mugundhan2017} used the striations in a type III radio burst
to determine the electron density variations along the path of the electron beams.
The observations of solar bursts in time, frequency and two spatial
coordinates naturally lead to 4D data.
Using the Murchison Widefield Array (MWA\footnote{See http://www.mwatelescope.org/.})
radio telescope data, \citet{Mohan2017} implement a formalism
to generate 4D data cubes based on brightness temperature maps.

\section*{Coronal Mass Ejections}
Coronal Mass Ejections (CMEs) are often associated with type II
and type IV radio bursts  observed over wide radio frequency range.
Simultaneous, radio and white-light observations of  CMEs
can be used to poorly constrained strength of the solar coronal magnetic field
above $\approx2$R$_{\odot}$ \citep{Kumari2017}.
Assuming plasma emission mechanism in type IV radio burst associated
with a behind-the-limb CME, \citet{Melnik2018} have estimated the densities
of plasma in the core of the CME. However, the relation between CMEs,
global EUV waves and type II solar is not always clear.
\citet{Long2017} studied over 160 global EIT waves observed in EUV
and found no clear relationship between global waves and type II radio bursts.
The relation between CMEs and type II is even more complicated
when the Sun launches multiple CMEs.
\citet{Al-Hamadani2017} studied type II solar radio bursts occurred
during a multiple coronal mass ejection event and
demonstrated that the last type II burst had enhanced emission
in a wider bandwidth, which should be consistent with the CME-CME interaction.
To understand the physics of the relation between the phenomena at the Sun
and the satellite damaging proton events at 1~AU, identification of solar flares  CMEs
responsible for the proton events is required.
The statistical relationships found by \citet{Miteva2018} serve
as a useful tool to diagnose the dependencies and test solar flare and CME models.

The CESRA 2016 workshop took place in Orl\'{e}ans, France. The
members of the Scientific Organizing Committee were M. Bart\`{a}
(Czech Republic), K.-L. Klein (France; co-chair), E.P. Kontar (UK),
M. Kretzschmar (France; co-chair), C. Marqu\'{e} (Belgium),
A. Nindos (Greece), S. Pohjolainen (Finland; co-chair), A. Warmuth
(Germany), and M.K. Georgoulis (Greece; as president of the European
Solar Physics Division). The workshop received financial support from the
University of Orl\'{e}ans, the Observatoire de Paris, and the CNRS/INSU.

The Special Issue Editors would like to thank both the authors
and the referees of the articles in this volume.


\end{article}

\end{document}